\documentclass[usenatbib,usegraphicx,twocolumn]{mn2e}
\newcommand{\e}{\end{equation}}
\newcommand{\bear}{\begin{eqnarray}}
\newcommand{\ear}{\end{eqnarray}}
\def\aj{AJ}
\def\apj{ApJ}
\def\jcap{JCAP}
\def\mnras{MNRAS}
\def\aap{A\&A}
\def\nat{Nature}
\def\apjs{ApJS}
\def\apjl{ApJ Letters}
\def\physrep {Physics Reports}

\begin{document}
\title[Cosmic homogeneity with Shannon entropy]{A method for testing
  the cosmic homogeneity with Shannon entropy}
    
\author[Pandey, B.]  {Biswajit Pandey$^{1,2}$\thanks{E-mail:
    biswa@mpa-garching.mpg.de}\\ 
  $^1$ Max-Planck Institute for
  Astrophysics, Karl-Schwarszchild Str. 1, D85748, Garching,
  Germany\\ 
  $^2$ Department of Physics, Visva-Bharati University,
  Santiniketan, Birbhum, 731235, India\\ }

       \maketitle
       
       \date{\today}
       
       \begin{abstract}
         
          We propose a method for testing Cosmic homogeneity based on
          the Shannon entropy in Information theory and test the
          potentials and limitations of the method on Monte Carlo
          simulations of some homogeneous and inhomogeneous 3D point
          process in a finite region of space. We analyze a set of
          N-body simulations to investigate the prospect of
          determining the scale of homogeneity with the proposed
          method and show that the method could serve as an efficient
          tool for the study of homogeneity.
         
       \end{abstract}

       \begin{keywords}
         methods: numerical - galaxies: statistics - cosmology: theory - large
         scale structure of the Universe.
       \end{keywords}
       
       \section{Introduction}

The cosmological principle which assumes that the Universe is
statistically homogeneous and isotropic on very large scales is one of
the fundamental pillars of modern cosmology. This can not be proved in
a mathematical sense and can be only verified from observations and
various predictions of the physical theories based on it. The cosmic
microwave background is by far the best conclusive evidence in favour
of isotropy \citep{penzias,smoot,fixsen} which also strongly supports
large scale homogeneity in the early Universe. Various other
observations like the isotropy in angular distributions of radio
sources \citep{blake} and the isotropy of the X-ray background
\citep{peeb93,wu,scharf} support the assumption of cosmic homogeneity
on large scales. The isotropy does not by itself guarantee homogeneity
and it implies homogeneity only when there is isotropy around every
points. The present Universe is known to be highly inhomogeneous on
small scales and there are important consequences if the
inhomogeneities persist on large scales. The most important
implication of inhomogeneities comes from the averaging problem in
General Relativity through their effect on the large scale dynamics
known as backreaction mechanism. The backreaction mechanism can cause
a global cosmic acceleration without any additional dark energy
component \citep{buchert97, schwarz, kolb06, buchert08} although it
seems unlikely that this can explain all of it
\citep{paranjape08}. The implications of inhomogeneities and
backreaction for Cosmology are still considered to be important even
if it does not provide an alternate explanation of dark energy
\citep{paranjape09, kolb10, ellis}.

The principle of cosmic homogeneity demands that the statistical
properties of the observed galaxy distribution in a given finite
volume does not depend on the location of that volume in the
Universe. The statistical properties of galaxy distributions are
characterized by the correlation functions \citep{peeb80}. The two
point correlation function on small scales $ 0.1 \, h^{-1} \rm {Mpc}
\leq r \leq 10 \, h^{-1} \rm {Mpc}$, is well described by a power law
of the form $\xi(r) = (\frac{r}{r_{0}})^{-\gamma}$, with correlation
length $r_{0} \sim 5 \, h^{-1} \rm {Mpc}$ and slope $\gamma \sim 1.8$.
$\xi(r)$ vanishes at scales $> 20 \, h^{-1} \rm {Mpc}$ which is
consistent with large scale homogeneity. However the problem with
correlation function analysis is that it assumes a mean density on the
scale of survey which is not a defined quantity below the scale of
homogeneity. Most of the statistical tools for homogeneity analysis
are based on the simple number counts $n(<r)$ in spheres of radius $r$
which is expected to scale as $\sim r^{3}$ for a homogeneous
distribution. The conditional density \citep{ hogg, labini11a}
measures the average density in these spheres which is expected to
flatten out beyond the scale of homogeneity. The fractal analysis
\citep{martinez90, coleman92, borgani95} uses the scaling of different
moments of $n(<r)$ to characterize the scale of homogeneity. Some of
the studies carried out with these methods on different galaxy surveys
claim to have found a transition to homogeneity on sufficiently large
scales $70-150 \, h^{-1} \rm {Mpc}$ \citep{martinez94, guzzo97,
  martinez98, bharad99, pan2000, kurokawa, hogg, yadav, prksh2,scrim}
whereas some studies claim the absence of any such transition out to
scale of the survey \citep{coleman92, amen, labini07, labini09a,
  labini09b, labini11b}. The disagreements between various studies
indicate the need for some alternative measures of homogeneity which
could capture interesting information on different aspects of a
homogeneous distribution and serve as complimentary and alternative
tool to the existing methods in the literature. In the present work we
introduce a method to asses homogeneity which is based on the
evenness, a more general and robust aspect of any homogeneous
distribution. We employ the Shannon entropy \citep{shannon48} to
measure the unevenness characterizing inhomogeneities in a
distribution and explore the possibility of employing the proposed
method for exploring the scale of homogeneity.
 
A brief outline of the paper follows. We describe our method and
various effects in Section 2, describe the tests and the data in
Section 3 and presents the results and conclusions in Section 4.
       
\section{Method of analysis}

Our method is based on the Shannon entropy in Information theory
originally proposed by Claude Shannon to quantify the information
content in strings of text.

In Information theory entropy is a measure of the amount of
information required to describe the random variable. The Shannon
entropy for a discrete random variable $X$ with $n$ outcomes
$\{x_{i}:i=1,....n\}$ is a measure of uncertainty denoted by $H(X)$
defined as,

\begin{eqnarray}
H(X) & = & - \sum^{n}_{i=1} \, p(x_i) \, log \, p(x_i)
\label{eq:shannon1}
\end{eqnarray}
where $p(x)$ is the probability distribution of the random variable
$X$. Increase in Shannon entropy increases the uncertainty and
decreases the information about the knowledge of the random
variable. Another interesting aspect of Shannon entropy is its
entropy-maximizing property for an uniform distribution.

We propose a method to study inhomogeneities in a 3D distribution of
points. Given a set of $N$ points distributed in 3D we consider each
of the $i^{th}$ points as center and determine $n_i(<r)$ the number of
other points within a sphere of radius $r$ as,\\
\begin{equation}
 n_i(<r)=\sum_{j=1}^{N}\Theta(r-\mid {\bf{x}}_i-\bf{x}_j \mid)
\label{eq:count}
\end{equation}
where $\Theta$ is the Heaviside step function and ${\bf{x}_{i}}$ and
${\bf{x}_{j}}$ are the radius vector of $i^{th}$ and $j^{th}$ points
respectively. To avoid any edge effects we discard all the points as
centers which lie within a distance $r$ from the survey
boundary. Clearly the number of valid centers will decrease with
increasing $r$ for any finite volume sample. We define a separate
random variable $X_{r}$ for each radius $r$ which has $M(r)$ possible
outcomes each given by, $f_{i,r}=\frac{\rho_{i,r}}{\sum^{M(r)}_{i=1}
  \, \rho_{i,r}}$ with the constraint $\sum^{M(r)}_{i=1} \,
f_{i,r}=1$. Here $\rho_{i,r}=\frac{n_{i}(<r)}{\frac{4}{3}\pi r^{3}}$
is the density at the $i^{th}$ center. Note that for a given sample
with a finite volume, $M(r)$ is the maximum numbers of valid centers
available at a radius $r$ i.e. there are no provision for projecting
further another sphere of radius $r$ within the given volume.

The Shannon entropy associated with the random variable $X_{r}$ can be
written as,
\begin{eqnarray}
H_{r}& = &- \sum^{M(r)}_{i=1} \, f_{i,r}\, log\, f_{i,r} \nonumber\\ &=& 
log(\sum^{M(r)}_{i=1}n_i(<r)) - \frac {\sum^{M(r)}_{i=1} \,
  n_i(<r) \, log(n_i(<r))}{\sum^{M(r)}_{i=1} \, n_i(<r)}
\label{eq:shannon2}
\end{eqnarray}
where the base of the logarithm is arbitrary and we choose it to be
$10$. Note that in $X_{r}$, $f_{r}$ and $H_{r}$, $r$ is a just a label
for a number and not an argument.

In an ideal situation when all the spheres around the $M(r)$ valid
centers are equally populated then one gets an uniform value of
$f_{i,r}=\frac{1}{M(r)}$ for all the centers maximizing the
uncertainty. Then the Shannon entropy $H_{r}$ (equation
3 ) has its maximum value $(H_{r})_{max}=log \, M(r)$
for radius $r$. The relative Shannon entropy
$\frac{H_{r}}{(H_{r})_{max}}$ at any $r$ quantifies the degree of
uncertainty in the knowledge of the random variable $X_{r}$. When
$\frac{H_{r}}{(H_{r})_{max}}=1$ the knowledge about the random
variable $X_{r}$ becomes most uncertain. The distribution of $f_{r}$
also become completely uniform when $\frac{H_{r}}{(H_{r})_{max}}=1$ is
reached. Equivalently one can use $1-\frac{H_{r}}{(H_{r})_{max}}$ to
quantify the information available in $X_{r}$ at any $r$.

The joint Shannon entropy of a set of independent random variables is
simply the sum of the individual entropies associated with each of the
random variables. But if they are dependent then the total entropy is
a sum of the conditional entropies. Entropy measures uncertainty in
the random variable whereas the information is a difference in
uncertainty that is a difference in entropies. The mutual information
characterizes the reduction in the uncertainty in the one random
variable due to the knowledge of the other. 

In our case the mutual information between the random variables
$X_{r}$ are always positive as the random variables are not
independent due to the fact that the density measurements around
$M(r)$ centers at each radius $r$ are taken from the same finite
volume sample. Given that the mutual information are always positive
for correlated random variables, an increase in information in $X_{r}$
at one $r$ would lead to decrease in information in $X_{r}$s at other
$r$ values. Apart from the correlations from finite sample, the random
variables could have extra correlations if the points are clustered or
if the points are distributed in a preferred way. These extra
correlations increase the mutual information in the random variables
$X_{r}$. The decrease of information in $X_{r}$ with increasing $r$ is
evident irrespective of a homogeneous/inhomogeneous distribution but
it would diminish differently depending on the nature and the degree
of inhomogeneity present in the distribution. Maximum uncertainty or
the complete loss of information in the knowledge of the random
variable $X_{r}$ at a radius $r$ suggests that beyond $r$ the random
variables $X_{r}$ would be independent and completely uninformative
about each other. In an infinite perfectly homogeneous system if the
set of spheres used for density measurements are completely
independent then the random variables $X_{r}$ at each $r$ has no
knowledge about each other making them maximally uncertain and devoid
of any information. When no inhomogeneities are present due to other
sources, the departure of $\frac{H_{r}}{(H_{r})_{max}}$ from $1$ would
be solely due to the correlations among $X_{r}$ caused by the finite
volume of the sample. Given the other sources are present one would
expect a larger departure of this ratio from $1$ and a more uneven
distribution of $f_{r}$. The scale where this departure levels up with
$1$ indicates absence of any correlations among $X_{r}$ beyond that
scale. But this ideal situation may never occur in an exact sense in a
finite sample as the correlations between $X_{r}$s introduced by the
finite volume and confinement bias persists over the whole range of
length scales. This would always give some residual information in
$X_{r}$ even at the largest value of $r$. But the mutual information
content due to clustering or any other source of inhomogeneity are
expected to diminish with increasing length scale provided a scale of
homogeneity exist for the distribution. As the exact transition marked
by $\frac{H_{r}}{(H_{r})_{max}}=1$ may never happen in a finite volume
sample so one could set a very small limiting value for
$1-\frac{H_{r}}{(H_{r})_{max}}$ to identify the scale of homogeneity.
In our analysis we set this limit to $10^{-4}$.

\subsection{Effects of clustering, intrinsic inhomogeneity and finite volume}

Clustering of the points is the most important source of correlations
between the random variables $X_{r}$. Clustering produces fluctuations
in $f_{r}$ which is directly related to the fluctuations in
$n(<r)$. When the points are clustered we expect on average
$\bar{n}(<r)$ number of points in a volume $V$ where $\bar{n}(<r)$ is
given by,\\
\begin{equation}
 \bar{n}(<r) = \lambda \, \int_{V} (1+\xi(x)) \, d^3x = \lambda \, V +
 4 \pi \lambda \int_{0}^{r} x^{2} \, \xi(x) \, dx
\label{eq:clustmean}
\end{equation}
 Here $\lambda$ is the mean density of the distribution and $V =
 \frac{4}{3} \pi r^{3}$ is the volume of the sphere used. The first
 term has fluctuations from the Poisson noise of the same order. The
 Poisson noise rapidly decreases with increasing $r$ and is only
 important on small scales. If the distribution itself is
 intrinsically inhomogeneous then Poisson noise will be modulated
 according to the spatially dependent intensity parameter $\lambda(x)$.
 The second term takes into account of clustering of the points. The
 variance of $n(<r)$ is,\\
\begin{equation}
\sigma_{n(<r)}^2 = \overline{n^2(<r)}-(\bar{n}(<r))^2
\label{eq:clustvar}
\end{equation}

Going back to our definition of
$f_{i,r}=\frac{\rho_{i,r}}{\sum^{M(r)}_{i=1} \, \rho_{i,r}}$, the
fluctuations in this quantity is closely related to the fluctuations
in $n_{i}(<r)$. The variance in $f_{r}$ is,

\begin{equation}
\sigma^{2}_{f_{r}} = \overline{f^2_{r}}-(\bar{f}_{r})^2 = \frac{\sigma_{n(<r)}^2}{\left[M(r) \, \bar{n}(<r)\right]^2}
\label{eq:fvar}
\end{equation}

which is $\frac{1}{M(r)^2}$ times the normalized variance of $n(<r)$.

One can quantify the correlations of any two random variables $X_{r}$
at two different $r$ by estimating their covariance. The correlation
coefficient is,
\begin{eqnarray}
C_{{X_{r_{i}}},{X_{r_{j}}}} & = & \frac{Cov \left(f_{r_{i}},f_{r_{j}}\right )}{\sqrt{\sigma^{2}_{f_{r_{i}}} \sigma^{2}_{f_{r_{j}}}}}
\label{eq:correlation}
\end{eqnarray}
 where the indices $i$ and $j$ takes value between $1$ to $n$. Here
 $n$ is the total number of $r$ values used in the analysis. In this
 case we have a positive correlation $0<C_{X_{r_{i}},X_{r_{j}}}<1$
 between $X_{r_{i}}$ and $X_{r_{j}}$ and the correlations would be
 higher when the differences between $r_{i}$ and $r_{j}$ are
 smaller. The random vector
 $X=\left(X_{r_{1}},X_{r_{2}},X_{r_{3}}....X_{r_{n}}\right)$ has
 $M(r_{1})M(r_{2})M(r_{3}).....M(r_{n})$ equally likely outcomes and
 in principle one can estimate the full covariance matrix of all the
 random variables $X_{r}$ at different $r$s to compute their
 correlations.

Even in the absence of any clustering and inhomogeneity one would
expect a positive correlations between the random variables as their
probability distribution is derived from the same finite volume
sample. The set of centers at each $r$ is a subset of the centers at
the preceding $r$. More specifically all the set of centers at each
$r$ is a subset of the set of the centers at the smallest $r$ implying
the spheres at different $r$ share some common regions. The fractional
amount of share is larger when the difference between $r$ is smaller
thereby introducing larger correlations between the $X_{r}$s from
neighboring $r$ values. So the finite volume introduces correlations
between the random variables $X_{r}$ at different $r$ which increases
the mutual information in the random variables.

\subsection{Effects of overlapping}

One should also keep in mind that the spheres used for density
measurements are not independent and could share large overlapping
regions. At each $r$ we use a finite set of spheres
$S_{r}=\{s_{1,r},s_{2,r},s_{3,r}.....,s_{M(r),r}\}$ for density
measurements. In general these spheres overlap with each other.  The
probability that a random point drawn out of the distribution would lie
in a particular sphere $s_{i,r}$ is $\frac{n_i(<r)}{N}$ where $N$ is
the total number of points in the distribution. The probability that
the randomly drawn point would appear somewhere in the sample is $1$.
If the spheres are disjoint then at any $r$ there are $M(r)+1$ outcome
of this experiment i.e the point would lie either in any one of the
$M(r)$ spheres or somewhere in the sample outside the spheres. But
given the fact that the spheres overlap the point could also appear at
the intersections of multiple spheres. Given a finite sample $A$ the
total probability can be written as,
\begin{eqnarray}
 P(A) &=& P(\cup_{i=1}^{M(r)} s_{i,r}) + P((\cup_{i=1}^{M(r)}
 s_{i,r})^{c}) \nonumber \\& & = \sum_{i=1}^{M(r)} P(s_{i,r}) -
 \sum_{i\neq j, 1}^{M(r)} P(s_{i,r}\cap s_{j,r}) \nonumber \\ & & 
+ \sum_{i\neq j \neq k, 1}^{M(r)} P(s_{i,r}\cap s_{j,r} \cap s_{k,r}) -... 
\nonumber \\ & &
 +\left(-1\right)^{M_{r}-1} P(\cap_{i=1}^{M(r)} s_{i,r}) + 
 P((\cup_{i=1}^{M(r)} s_{i,r})^{c}) \nonumber \\ & & =
 \sum_{i=1}^{M(r)} P(s_{i,r}) - P_{overlap} +P((\cup_{i=1}^{M(r)}
 s_{i,r})^{c}) \nonumber \\ & & = 1
\label{eq:overlap}
\end{eqnarray}
The first sum in above equation gives the sum of probabilities for
individual spheres provided they are disjoint. As the spheres overlap
one needs to take into account the probabilities for all the
possibilities. The second sum is over all distinct pairs of spheres,
third sum is over all distinct triples of spheres and so forth.  The
last term which gives the probability that the point would come up in a
region $(s_{1,r}\cup s_{2,r}\cup s_{3,r}...\cup s_{M(r),r})^{c}$
outside the union of spheres is important only at small $r$ and
becomes insignificant afterwards. It may be noted that in an
overlapping scenario the different terms in equation \ref{eq:overlap}
can be much larger than $1$ but when summed together they will always
give $1$. In equation \ref{eq:overlap}, $P_{overlap}$ contains all the
terms from overlapping spheres and successive terms in $P_{overlap}$
has alternate sign . At smaller $r$ the most dominant contribution to
$P_{overlap}$ comes from all pairs of overlapping spheres. The
contribution from all triples of overlapping spheres would be the next
dominant term and so on. Thus at smaller $r$ all the individual sums
in $P_{overlap}$ are different and the magnitude of each sum is lesser
than its preceding sum resulting into a positive value of
$P_{overlap}$. New sums appear in $P_{overlap}$ with increasing $r$
thereby increasing its value. But at the same time the differences
between the successive sums starts decreasing due to larger overlap
thereby decreasing the value of $P_{overlap}$. There is a competition
between this two effect and initially the first effect is more
dominant than the second leading to an overall increase in
$P_{overlap}$ with increasing $r$. As we increase the radius $r$ the
numbers of valid centers $M(r)$ decreases and they preferentially get
more confined near the center of the sample. Due to the finite volume
of the sample ultimately the second effect will dominate on some scale
depending on the size of the finite sample and also depending on the
nature of inhomogeneity to some extent. For example the confinement
bias is expected to be even higher for an inhomogeneous distribution
which preferentially has more particles residing near the center of
the sample. When the differences between the successive sums become
smaller then there would be a larger net cancellation leading to
smaller values of $P_{overlap}$. As the second effect starts
dominating the first one, the value of $P_{overlap}$ would start
decreasing and finally on the scale of the largest sphere that would
fit inside the finite sample, all the individual terms in all the
individual sums in $P_{overlapp}$ would be of the same order
i.e. $P(s_{i})\sim P(s_{i}\cap s_{j}) \sim P(s_{i}\cap s_{j} \cap
s_{k}) \sim P(s_{1}\cap s_{2} \cap s_{3}......\cap s_{M(r)}) \sim
1$. This would lead to a very large mutual cancellation of all the
sums in $P_{overlapp}$ decreasing its value to $P(\cup_{i=1}^{M(r)}
s_{i,r})-1$ or $M(r)-1$. This finite volume effect eventually
introduces an artificial evenness in the distribution of $f_{r}$ at
some $r$ depending on the sample size.  We see that the effect of
overlapping starts dominating much before the scale of the largest
possible sphere (Figure \ref{fig:ovp}) making an interpretation difficult
on large scales. We note that all the other methods of testing
homogeneity based on count in spheres $n(<r)$ are also affected by the
same problem making the interpretations on large scales equally
difficult.

In order to avoid these complicacy of correlations introduced by the
overlapping of the spheres we also consider non-overlapping spheres of
different radii. But the statistics becomes too noisy due to very
small number of independent spheres at progressively larger radii
which consequently prohibits us to address the issue of homogeneity on
large scales.

A point of caution in this context is that in our method
$\frac{H_{r}}{H_{{r}_{max}}} \sim 1$ at very large length scales does
not necessarily indicate a real transition to homogeneity which is an
obvious outcome forced by the confinement bias resulting from the
finite volume of the sample. But with a large enough sample which
ensures spheres upto a sufficiently large $r$ without significant
confinement of the centers could detect the scale of homogeneity if
the transition happens before the scale where the confinement bias
completely dominates the statistics.

\section{Tests on homogeneous, inhomogeneous and clustered distributions }

In order to study the prospects and limitations of the proposed method
we carry out some preliminary tests by applying it to some simple
distributions. We consider (i) homogeneous distribution without any
clustering, (ii) inhomogeneous distribution without any clustering and
(iii) strongly clustered distribution.

In above three cases the unevenness in type (i) is only due to Poisson
noise. The unevenness in type (ii) distribution is controlled by
Poisson noise too but it has the additional complexity that the
contribution to the unevenness is governed by the distribution of the
spatially dependent intensity parameter $\lambda(x)$. In type (iii)
distribution both Poisson noise and the clustering together
contributes to the unevenness.
 
For the first two types we generate a set of Monte Carlo realizations
of some simple homogeneous and inhomogeneous point process.  For the
third type we use a set of N-body simulations where the points are
strongly clustered.

\subsection{Monte Carlo simulations of homogeneous and inhomogeneous point 
processes}

 We generate a set of Monte Carlo realizations of different types of
 homogeneous and inhomogeneous distributions.

For the sake of simplicity we consider some simple radial density
distributions $\rho(r) = K \, \lambda(r)$ where (i) $\lambda(r) =
\frac{1}{r}$ , (ii) $\lambda(r) = \frac{1}{r^{2}}$ and (iii)
$\lambda(r) = 1$. $K$ is a normalization constant. The distributions
in (i) and (ii) are inhomogeneous Poisson distributions which are
isotropic about only one point and the inhomogeneities in these
distributions persist on all scales. The distribution in (iii) is a
homogeneous Poisson point process which has a constant density
everywhere.

Enforcing the desired number of particles $N$ within radius $R$ one
can turn the radial density function into a probability function within
$r=0$ to $r=R$ which is normalized to one when integrated over that
interval. So the probability of finding a particle at a given radius
$r$ is $P(r)=\frac{ r^2 \lambda(r) } {\int_{0}^{R} r^2 \lambda(r) \,
  dr}$ which is proportional to the density at that radius implying
more particles in high density regions.

We generate the Monte Carlo realizations of these distributions using
a Monte Carlo dartboard technique. The maxima of the function $r^2
\lambda(r)$ in $P(r)$ is at $r=R$ in (i) and (ii) whereas in (iii) it
is same and constant everywhere. We label the maximum value of $P(r)$
as $P_{max}$. We randomly choose a radius $r$ in the range $0 \le r
\le R$ and a probability value is randomly chosen in the range $0 \le
P(x) \le P_{max}$. The actual probability of finding a particle at the
selected radius is then calculated using expression for $P(r)$ and
compared to the randomly selected probability value. If the random
probability is less than the calculated value, the radius is accepted
and assigned isotropically selected angular co-ordinates $\theta$ and
$\phi$, otherwise the radius is discarded. In this way, radii at which
particle is more likely to be found will be selected more often
because the random probability will be more frequently less than the
calculated actual probability. We choose $R=200$ in $h^{-1} \, {\rm
  Mpc}$ unit and $N=10^{5}$. We generate $10$ realizations of each of
the above density distributions and analyze them separately with the
method described in section 2.

\subsection{N-body simulations}
We simulate the dark matter distribution using a Particle-Mesh (PM)
N-body code. The simulations use $256^3$ particles on a $512^3$ mesh
and cover a comoving volume of $[921.6 h^{-1} {\rm Mpc}]^3$. We use
$(\Omega_{m0},\Omega_{\Lambda0},h)=(0.27,0.73,0.71)$ for the
cosmological parameters along with a $\Lambda$CDM power spectrum with
spectral index $n_s=0.96$ and normalization $\sigma_8=0.812$
(\citealt{komatsu}). A simple ``sharp cutoff'' biasing, scheme
\citep{cole} was used to extract particles from the simulations that
are biased relative to the dark matter and are labelled as
galaxies. The bias parameter $b$ of each simulated biased sample was
estimated using the ratio
\begin{equation}
b=\sqrt{ \frac {\xi_g(r)}{\xi_{dm}(r)}}
\end{equation}
where $\xi_g(r)$ and $\xi_{dm}(r)$ are the galaxy and dark matter
two-point correlation functions respectively. This ratio is found to
be constant at length-scales $r \ge 5 h^{-1} {\rm Mpc}$ and we use the
average value over $5-40 h^{-1}{\rm Mpc}$. We use this method to
generate galaxy samples with bias values $b=1.5,2$ and $2.5$. We run
the simulation for three different realizations of the initial density
fluctuations and extract biased samples from each of them. We extract
randomly $N=10^5$ particles from three non overlapping spherical
regions of radius $R=200 \, h^{-1} {\rm Mpc}$ from each of the
simulation boxes giving us total nine samples for each of the bias
values. The numbers $N$ and $R$ and the specific spherical geometry
are chosen just to maintain the uniformity in all the analysis
presented here. Generally one can choose any number of particles and
any geometry for the samples.

\begin{figure}
\rotatebox{-90.0} {\scalebox{0.3}{\includegraphics{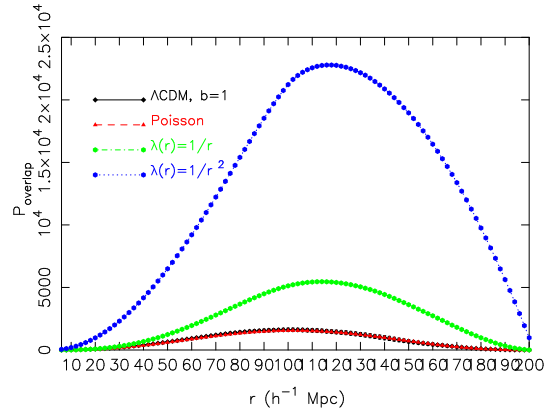}}}
\caption{This shows the $P_{overlap}$ (equation \ref{eq:overlap}) as a
  function of $r$ for different distributions as indicated in the
  figure. The error bars overplotted on the data points here are not
  visible due to their very small sizes.}
  \label{fig:ovp}
\end{figure}

\begin{figure}
\rotatebox{-90.0} {\scalebox{0.3}{\includegraphics{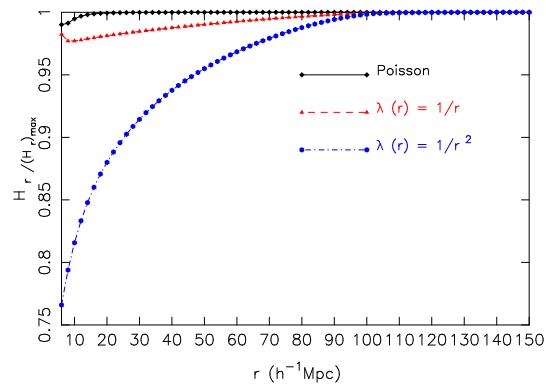}}}
\caption{ This shows the ratio $\frac{H_{r}}{(H_{r})_{max}}$ as a
  function of $r$ for homogeneous Poisson distribution and two
  different anisotropic distributions. The tiny error bars overplotted
  on the data points are invisible here. }
  \label{fig:shratio1}
\end{figure}

\begin{figure}
\rotatebox{-90.0} {\scalebox{0.3}{\includegraphics{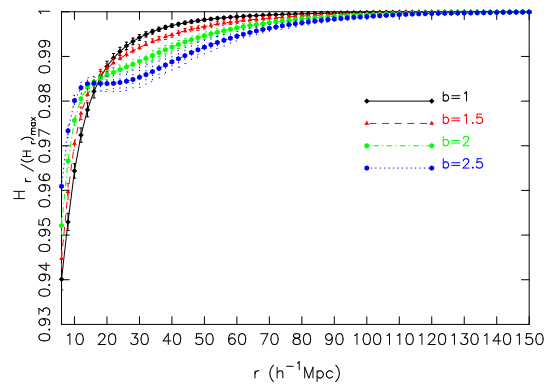}}}
\caption{ This shows the ratio $\frac{H_{r}}{(H_{r})_{max}}$ as a
  function of $r$ for the unbiased $\Lambda$CDM model and its
  different biased variants. $1-\sigma$ error bars obtained from $9$
  different samples are overplotted at each points.}
  \label{fig:shratio2}
\end{figure}

\section{Results and Conclusions}

 In the Figure \ref{fig:shratio1} we show the variations of
 $\frac{H_{r}}{(H_{r})_{max}}$ with distance $r$ for some homogeneous
 and inhomogeneous distributions described in subsection 3.1. For the
 Poisson distribution the value of $\frac{H_{r}}{(H_{r})_{max}}$ shows
 a small departure from $1$ at smaller values of $r$ which decreases
 with increasing $r$. The ratio reaches a value $ \sim 1$ at $r \sim
 15$. As the points are uncorrelated, the Void Probability function
 \citep{sdm} for a homogeneous Poisson distribution is $e^{-\lambda \,
   V}$. It is expected that such a distribution would become
 homogeneous on a scale $r \sim {\lambda}^{-\frac{1}{3}}$, which is a
 measure of the average size of voids in the distribution. Since the
 correlations between the random variables $X_{r}$ in the homogeneous
 Poisson distribution mainly comes from the finite volume effect one
 would expect a lesser amount of mutual information in the set of
 random variables and consequently a faster information leakage but
 the information in $X_{r}$ i.e. $1-\frac{H_{r}}{(H_{r})_{max}}$ would
 never be exactly zero due to persistence of the finite volume
 correlations at all $r$.

The $\frac{H_{r}}{(H_{r})_{max}}$ values at smaller $r$ values shows
relatively larger departure from $1$ for both the inhomogeneous
distributions considered here. The ratio has a larger departure for
the distribution with $\lambda(r)=\frac{1}{r^{2}}$ than the
distribution with $\lambda(r)=\frac{1}{r}$ on all scales. This shows
that the ratio $\frac{H_{r}}{(H_{r})_{max}}$ describes the degree of
inhomogeneity present in two different anisotropic distributions in
correct order. Both these distributions are only isotropic about one
point and have a radial density distribution which does not allow them
to be homogeneous on any scales. The ratio
$\frac{H_{r}}{(H_{r})_{max}} \sim 1$ at $r \sim 120$ for both of them
simply shows the importance of confinement bias which ultimately is
expected to take over the situation at some $r$ depending on the
sample size and type of inhomogeneity present. The confinement bias
gets a larger boost in both of these distributions (Figure
\ref{fig:ovp}) as there are already more particles preferentially
located near the center of the samples. This effect produces larger
overlap and hence larger mutual information are stored in the random
variables from neighbouring $r$s. Consequently this accelerates the
leakage of information in $X_{r}$ with increasing $r$ finally forcing
the ratio $\frac{H_{r}}{(H_{r})_{max}} \sim 1$. It may be noted that
for the homogeneous Poisson distribution there are no such additional
boost coming from preferential deposition and further there are no
extra correlations due to clustering. The transition
$\frac{H_{r}}{(H_{r})_{max}} \sim 1$ is reached there at a $r$ much
before the scale where the confinement bias play a dominant role
indicating a real transition to homogeneity.
  
 In the Figure \ref{fig:shratio2} the variations of
 $\frac{H_{r}}{(H_{r})_{max}}$ with distance $r$ are shown for
 $\Lambda$CDM N-body simulations with different linear bias
 values. The assumption of linear bias holds reasonably well on large
 scales. Different types of galaxies are biased differently with
 respect to the dark matter and the inhomogeneities in the
 distributions of different types of galaxies could be modulated by
 their bias values. We would also like to investigate here how well
 the ratio $\frac{H_{r}}{(H_{r})_{max}}$ can track the variation in
 inhomogeneities in a distribution on different scales and give some
 useful information about the characteristics of the inhomogeneities
 present in them. We see that the ratio initially depart from $1$ at
 small $r$ for all the distributions with higher bias values showing
 systematically lesser inhomogeneity (lesser information in $X_{r}$)
 than lower bias values. In the unbiased $\Lambda$CDM model the
 centers used for calculating $n(<r)$ are residing in all types of
 environments (clusters, sheets, filaments and voids). The fact that
 the centers are distributed across different types of nonlinear
 structures increases the unevenness in the distribution of $f_{r}$
 and the mutual information in $X_{r}$. Whereas with increasing bias
 values the particles in a biased distribution preferentially
 represent progressively higher density peaks in the density field
 homogenizing the distribution on the corresponding length
 scales. This is primarily due to the fact that in a biased
 distribution the centers are located in less diverse environments. A
 complete reversal of this behaviour is seen at $15-20 \, h^{-1} \rm
 {Mpc}$ after which with increasing bias values the distributions
 systematically show larger inhomogeneity and more information in
 $X_{r}$ at all scales $r$ until they eventually merge to
 $\frac{H_{r}}{(H_{r})_{max}}=1$. This behaviour is related to the
 fact that increasing the radius $r$ beyond the typical scale of the
 nonlinear structures in the biased distributions would lead to larger
 disparity in $n(<r)$ and $f_{r}$ values and hence the larger
 inhomogeneities. The scales corresponding to the reversal in the
 behaviour of ratio corresponds to the typical scale of the nonlinear
 structures present in the biased distributions. Further on large
 scales the correlation functions $\xi(r)$ in biased distributions are
 $b^{2}$ times larger than the $\xi(r)$ in unbiased distributions
 introducing larger correlations among all the random variables
 $X_{r}$ at larger $r$. Eventually all the curves merge to
 $\frac{H_{r}}{(H_{r})_{max}}=1$ in the range $100-150 \, h^{-1} \rm
 {Mpc}$ with the difference that the transition appears to happen at a
 relatively larger $r$ values for larger bias. Although this can not
 be emphasized due to the size of the overlapping error bars at those
 length scales. It may be noted that the inhomogeneities in the
 $\Lambda$CDM model with different biases are much smaller as compared
 to the inhomogeneous radial density distributions given by
 $\lambda(r)=\frac{1}{r^2}$. The information in $X_{r}$ for a Poisson
 distribution mostly comes from the finite volume effect whereas the
 very large mutual information in the random variables in two
 inhomogeneous distributions is the combined outcome of finite volume
 correlations and large degree of confinement bias. In the unbiased
 and biased $\Lambda$CDM models the mutual information are generated
 by correlations from confinement bias and clustering together. It may
 be noted that the confinement biases are much smaller and very
 similar in the $\Lambda$CDM models and the homogeneous Poisson
 distributions (Figure \ref{fig:ovp}). The confinement biases become
 very large in the two anisotropic distributions considered here which
 eventually force the ratio $\frac{H_{r}}{(H_{r})_{max}}$ to $1$ at
 some scales.

If a scale of homogeneity exists then ideally one would expect a
naturally emerging evenness in $f_{r}$ when the spheres of that radius
around each centers statistically include similar numbers of different
types of nonlinear structures. So provided that the correlations
introduced by the finite volume of the sample and the confinement bias
are much smaller compared to the correlations induced by clustering
then the leakage of information in $X_{r}$ would ideally track the
variation of inhomogeneity in the distribution with scale $r$. But
unfortunately in a real situation one has to also deal with the
correlations due to finite volume and confinement bias modulating the
information in $X_{r}$ or the evenness of $f_{r}$. We find that the
confinement bias could start to dominate even much before the scale
defined by the largest sphere included in any finite sample (Figure
\ref{fig:ovp}). The degree of correlations introduced among $X_{r}$s
would also depend on the type of inhomogeneity present in the
distribution specially if it is overpopulated near the center of the
sample with respect to the rest of the volume resulting into larger
overlap between the spheres. When dealing with galaxy redshift surveys
one has to also properly take into account the different selection
effects involved, redshift space distortions and the specific geometry
of the samples. We plan to carry out analysis in the publicly
available catalogues from large galaxy redshift surveys (e.g. 2dFGRS,
\citealt{colles}; SDSS, \citealt{stout}; 2MASS, \citealt{huch}) in
future works and investigate these issues further. It may be noted
here that a caveat of our method is we assume that we always have
access to the data samples on a spatial hypersurface of constant
time. Though this assumption is approximately valid in case of low
redshift galaxy samples in the nearby Universe or data samples from a
snapshot of N-body simulations, it is not strictly true as the entire
observational samples do not consist of objects on a constant time
hypersurface but rather on a light cone. Taking this into account it
is hard to distinguish radial inhomogeneity from time evolution
without assuming a cosmological model. Consequently this could make an
inhomogeneous but isotropic model (e.g. anti-Copernican void models)
to look like a homogeneous one on large scales. Fortunately these
models can be constrained with other observations such as SNe, CMB,
BAO and measurements of Hubble parameter \citep{zibin, clifton,
  biswas, chris}.

Finally we note that the method presented here has the desired ability
to characterize inhomogeneities and their variations on different
length scales in any 3D distribution of points. The method has the
potential to successfully identify any existing scale of homogeneity
given a sufficiently large volume which can ensure negligible
confinement bias and less effective overlap between the spheres upto
large length scales. Given the large survey volumes that are currently
available from the modern redshift surveys the method could provide an
efficient tool for exploring the issue of Cosmic homogeneity.

\section*{Acknowledgment}
The author thanks Somnath Bharadwaj for useful comments and
discussions. The author would also like to thank an anonymous refree
for insightful comments about the paper. The author acknowledges the
Alexander von Humboldt Foundation for financial support and the
Computing Center of the Max Planck Society in Garching (RZG) for the
computing facilities provided for the work.

\end{document}